\documentclass[12pt]{article}
\usepackage[english]{babel}%
\usepackage{graphicx}
\usepackage{amsfonts}

\newcommand{\sgn}{{\bf sgn}}

\begin{document}

\newtheorem{theorem}{Theorem}
\newtheorem{example}{Example}
\newtheorem{tests}{Tests}

\renewcommand{\thesection}{\arabic{section}}
\renewcommand{\thetheorem}{\arabic{section}.\arabic{theorem}}
\renewcommand{\theequation}{\arabic{section}.\arabic{equation}}
\renewcommand{\theexample}{\arabic{section}.\arabic{example}}

\def\qed{{\bf \hfill $\Box$}\endtrivlist}

\begin{center}{\Large Analytical Solution for the Generalized Fermat-Torricelli Problem} \end{center}

\begin{center}{\large Alexei Yu. Uteshev\footnote{alexeiuteshev@gmail.com}}  \end{center}

\begin{center}{\it St. Petersburg State University, St. Petersburg, Russia} \end{center}


We present explicit analytical solution for  the problem of minimization of the function
$ F(x,y)= \sum_{j=1}^3 m_j \sqrt{(x-x_j)^2+(y-y_j)^2} $, i.e. we find the coordinates
of stationary point and the corresponding critical value as functions of $ \{m_j,x_j,y_j \}_{j=1}^3 $.
In addition, we also discuss inverse problem of finding such values for $ m_1,m_2,m_3 $ for which the
corresponding function $ F $ possesses a prescribed position of stationary point.


\section{Introduction}
\setcounter{equation}{0}
\setcounter{theorem}{0}
\setcounter{example}{0}

Given the coordinates of three noncollinear points $ P_1=(x_1,y_1),P_2=(x_2,y_2),P_3=(x_3,y_3) $ in the plane,
find the coordinates of the point $ P_{\ast}=(x_{\ast},y_{\ast}) $ which gives a solution for the optimization problem
\begin{equation}
 \min_{(x,y)} F(x,y) \quad \mbox{ for } \quad  F(x,y)= \sum_{j=1}^3m_j \sqrt{(x-x_j)^2+(y-y_j)^2} \ .
 \label{F}
 \end{equation}
Here  $ m_1,m_2,m_3 $ are assumed to be real positive numbers and will be subsequently referred to as \emph{weights}.

The stated problem in its particular case of equal weights $ m_1=m_2=m_3=1 $ is known since 1643 as (classical) Fermat-Torricelli problem.
It has a unique solution which coincides either with one of the point $ P_1,P_2,P_3 $ or with the so-called \emph{Fermat} or \emph{Fermat-Torricelli}
\emph{point} \cite{b1,b10} of the triangle $ P_1P_2P_3 $.

Generalization of the problem to the case of unequal weights was investigated since XIX century.
This generalization is known under different names:  Steiner problem, Weber problem,  problem of railway junction\footnote{(\emph{Germ.}) Problem des Knotenpunktes} \cite{b2,b4}, the three factory problem \cite{b3}.
The two last names were inspired by the optimal transportation problem like a following one. Let the cities $ P_1,P_2 $ and $ P_3 $ be the sources of iron ore, coal and water
respectively. In order to produce one tonne of steel one the steel works need $ m_1 $ tonnes of iron, $ m_2 $ tonnes of coal and $ m_3 $ tonnes of water. Assuming that the freight charge
for tonne-kilometer is independent of the nature of the cargo, find the optimal position for steel works connected with $ P_1,P_2,P_3 $ via straight roads so as to
minimize the transport costs.

In the rest of the paper this problem will be referred to as the \emph{generalized Fermat-Torricelli problem}. Existence and uniqueness of its solution is guaranteed by the following
result \cite{b10}:

\begin{theorem}  \label{Th1} Denote by $ \alpha_1, \alpha_2,\alpha_3 $ the corner angles of the triangle $ P_1P_2P_3 $.
If the conditions
\begin{equation}
\left\{
\begin{array}{ccc}
m_1^2&<&m_{2}^2+m_{3}^2+2\,m_2m_3 \cos \alpha_1 , \\
m_2^2&<&m_{1}^2+m_{3}^2+2\,m_1m_3 \cos \alpha_2 ,   \\
m_3^2&<&m_{1}^2+m_{2}^2+2\,m_1m_2 \cos \alpha_3
\end{array}
\right.
\label{r123}
\end{equation}
are fulfilled then there exists a unique solution $ P_{\ast}=(x_{\ast},y_{\ast}) \in \mathbb R^2 $ for the generalized Fermat-Torricelli problem lying inside
the triangle $ P_1P_2P_3 $. This point is a stationary point for the function $ F(x,y) $, i.e. a real solution of the system
\begin{equation}
\sum_{j=1}^3 \frac{m_j(x-x_j)}{\sqrt{(x-x_j)^2+(y-y_j)^2}}=0,\quad
\sum_{j=1}^3 \frac{m_j(y-y_j)}{\sqrt{(x-x_j)^2+(y-y_j)^2}}=0 \ .
\label{sys1}
\end{equation}
If any of the conditions (\ref{r123}) is violated then $ F(x,y) $ attains its minimum value at the corresponding vertex of the triangle.
\end{theorem}

Let us overview some approaches for finding the point $ P_{\ast} $. The first one is geometrical: the point is
found as intersection point of special construction of lines or circles. For the equal weighted case, Torricelli proved that
the circles circumscribing the equilateral triangles constructed on the sides of
and outside the triangle $ P_1P_2P_3 $ intersect in the point $ P_{\ast} $; for an alternative Simpson construction of $ P_{\ast} $ see \cite{b11}. For the general (unequal weighted) case see \cite{b2,b4}.

The second approach is based on mechanical model\footnote{Sometimes incorrectly called P\'olya's mechanical model.}: a horizontal board is drilled with the
holes at the points $ P_1,P_2,P_3 $ (or at the vertices of a triangle similar to $ P_1P_2P_3 $) ; three strings are tied together in a knot
at one end, the loose ends are passed through the holes and are attached to physical weights proportional to $ m_1,m_2, m_3 $ respectively below the board.
The equilibrium position of the knot yields the solution \cite{b2}.

The third approach, based on gradient descent method, was originated in the paper \cite{b13}; further developments and comments can be found in \cite{b14,b15}.

The present paper is devoted to the fourth approach --- analytical one. We look for explicit expressions for the coordinates of
stationary point  $ P_{\ast} $ as functions of $ \{m_j,x_j,y_j\}_{j=1}^3 $.  Although the existence of such a solution by radicals, i.e. in a
finite number of operations like standard arithmetic ones and extraction of  (positive integer) roots, is not questioned in any
review article on the problem, we failed to find in publications the constructive and universal version of an algorithm even for the
classical (i.e. equal weighted) case.

\section{Algebra}
\setcounter{equation}{0}
\setcounter{theorem}{0}
\setcounter{example}{0}

\begin{theorem} \label{Th2} Under the conditions (\ref{r123}), the coordinates of stationary point $ (x_{\ast},y_{\ast}) $ of the function $ F(x,y) $ are as follows:
\begin{equation}
x_{\ast}=\frac{K_1K_2K_3}{4 S \sigma d} \left(\frac{x_1}{K_1}+\frac{x_2}{K_2}+\frac{x_3}{K_3} \right), \
y_{\ast}=\frac{K_1K_2K_3}{4 S \sigma d} \left(\frac{y_1}{K_1}+\frac{y_2}{K_2}+\frac{y_3}{K_3} \right)
\label{GFTp}
\end{equation}
with
$$
F(x_{\ast},x_{\ast})=\min_{(x,y)} F(x,y)=\sqrt{d} \ .
$$
Here
\begin{equation}
d=\frac{1}{2\sigma} (m_1^2K_1+m_2^2K_2+m_3^2K_3)
\label{d_1}
\end{equation}
\begin{equation}
= 2\, S \sigma + \frac{1}{2}\left[m_1^2(r_{12}^2+r_{13}^2-r_{23}^2)+
m_2^2(r_{23}^2+r_{12}^2-r_{13}^2)+m_3^2(r_{13}^2+r_{23}^2-r_{12}^2) \right] \ .
\label{d_2}
\end{equation}
and
$$
r_{j\ell}=\sqrt{(x_j-x_{\ell})^2+(y_j-y_{\ell})^2}=|P_jP_{\ell}| \quad for \ \{j,\ell\} \subset \{1,2,3\} \ ;
$$
\begin{equation}
S=|x_1y_2+x_2y_3+x_3y_1-x_1y_3-x_3y_2-x_2y_1| \ ;
\label{S}
\end{equation}
\begin{equation}
\sigma= \frac{1}{2} \sqrt{-m_1^4-m_2^4-m_3^4+2\,m_1^2m_2^2+2\,m_1^2m_3^2+2\,m_2^2m_3^2} \ ;
\label{sigma}
\end{equation}
and
\begin{equation}
\left\{
\begin{array}{ccc}
K_1&=&(r_{12}^2+r_{13}^2-r_{23}^2)\sigma+(m_2^2+m_3^2-m_1^2) S , \\
K_2&=&(r_{23}^2+r_{12}^2-r_{13}^2)\sigma+(m_1^2+m_3^2-m_2^2) S , \\
K_3&=&(r_{13}^2+r_{23}^2-r_{12}^2)\sigma+(m_1^2+m_2^2-m_3^2) S .
\end{array}
\right.
\label{K1K2K3}
\end{equation}
\end{theorem}

\textbf{Proof.} Firstly, we established via direct computations the validity of the following  equalities:
\begin{equation}
K_1K_2+K_1K_3+K_2K_3 = 4\sigma S d  \ ,
\label{dd2}
\end{equation}
and a dual one for (\ref{d_1}):
\begin{equation}
r_{23}^2K_1+r_{13}^2K_2+r_{12}^2K_3= 2\, S d \ .
\label{dd3}
\end{equation}

Secondly, let us deduce the following relationships:
\begin{equation}
\sqrt{(x_{\ast} -x_j)^2+(y_{\ast} -y_j)^2} = \frac{m_jK_j}{2\sigma \sqrt{d}}  \quad for \ j \in \{1,2,3\} \ .
\label{dd1}
\end{equation}
To prove (\ref{dd1}) for $ j=1 $ we first represent $ x_{\ast} $ and $ y_{\ast} $ given by (\ref{GFTp}) with the aid of (\ref{dd2}):
\begin{equation}
x_{\ast}=\frac{1}{\frac{1}{K_1}+\frac{1}{K_2}+\frac{1}{K_3}} \left(\frac{x_1}{K_1}+\frac{x_2}{K_2}+\frac{x_3}{K_3} \right),\
y_{\ast}=\frac{1}{\frac{1}{K_1}+\frac{1}{K_2}+\frac{1}{K_3}} \left(\frac{y_1}{K_1}+\frac{y_2}{K_2}+\frac{y_3}{K_3} \right).
\label{dd4}
\end{equation}
Thus,
$$
(x_{\ast} -x_1)^2+(y_{\ast} -y_1)^2=\left(\frac{K_1K_2K_3}{4\,\sigma S d}\right)^2 \left[ \left(\frac{x_2}{K_2}+ \frac{x_3}{K_3} - \frac{x_1}{K_2}-\frac{x_1}{K_3} \right)^2 +
\left(\frac{y_2}{K_2}+ \frac{y_3}{K_3} - \frac{y_1}{K_2}- \frac{y_1}{K_3} \right)^2 \right]
$$
$$
=\left(\frac{K_1K_2K_3}{4\,\sigma S d}\right)^2\bigg[ \frac{(x_2-x_1)^2+(y_2-y_1)^2}{K_2^2} + \frac{(x_3-x_1)^2+(y_3-y_1)^2}{K_3^2}
$$
$$
\qquad \qquad \qquad +2\, \frac{(x_2-x_1)(x_3-x_1)+(y_2-y_1)(y_3-y_1)}{K_2K_3} \bigg]
$$
$$
=\left(\frac{K_1K_2K_3}{4\,\sigma S d}\right)^2 \left[\frac{r_{12}^2}{K_2^2}+\frac{r_{13}^2}{K_3^2} +2 \frac{1/2(r_{12}^2+r_{13}^2-r_{23}^2)}{K_2K_3} \right]
$$
$$
=\frac{K_1^2}{(4\,\sigma S d)^2} \left[r_{12}^2K_3^2+r_{13}^2K_2^2+(r_{12}^2+r_{13}^2-r_{23}^2)K_2K_3 \right]
$$
$$
=\frac{K_1^2}{(4\,\sigma S d)^2} \left[(r_{12}^2K_3+r_{13}^2K_2)(K_2+K_3)-r_{23}^2K_2K_3 \right]
$$
$$
\stackrel{(\ref{dd3})}{=}
\frac{K_1^2}{(4\,\sigma S d)^2} \left[(2\,S d-r_{23}^2K_1)(K_2+K_3)-r_{23}^2K_2K_3 \right]
$$
$$
=\frac{K_1^2}{(4\,\sigma S d)^2} \left[2\,S d(K_2+K_3)-r_{23}^2(K_1K_2+K_1K_3+K_2K_3) \right]
$$
$$
\stackrel{(\ref{dd2})}{=}
\frac{K_1^2}{(4\,\sigma S d)^2} \left[2\,S d(K_2+K_3)-4\,r_{23}^2 \sigma S d \right] =
\frac{2\,S d K_1^2}{(4\,\sigma S d)^2} \left[K_2+K_3-2\,r_{23}^2 \sigma \right]
$$
$$
\stackrel{(\ref{K1K2K3})}{=} \frac{K_1^2}{8\,Sd \sigma^2 } \left[2\,m_1^2S \right]= \frac{m_1^2K_1^2}{4\sigma^2 d} \ .
$$
Similar arguments hold for $ j \in \{2,3\} $ in (\ref{dd1}). To complete the proof of those equalities it should be additionally verified that the values $ K_1,K_2,K_3 $ are nonnegative. This will be done in the next section.

Now let us prove the first statement of the theorem.  Substitute (\ref{GFTp}) into the left-hand side of the first equation of (\ref{sys1}). The resulting expression can be represented with the aid of formulae (\ref{dd2}) and  (\ref{dd1}) as
$$
\frac{x_{\ast}-x_1}{K_1}+\frac{x_{\ast}-x_2}{K_2}+\frac{x_{\ast}-x_3}{K_3}=x_{\ast} \left(\frac{1}{K_1}+\frac{1}{K_2}+\frac{1}{K_3}\right)-
\left(\frac{x_1}{K_1}+\frac{x_2}{K_2}+\frac{x_3}{K_3}\right)\stackrel{(\ref{dd4})}{=}0 .
$$
Similar arguments are valid for the second equation from (\ref{sys1}). Finally compute  $ F(x_{\ast},y_{\ast}) $:
$$
F(x_{\ast},y_{\ast})=\sum_{j=1}^3 m_j \sqrt{(x_{\ast} -x_j)^2+(y_{\ast} -y_j)^2} \stackrel{(\ref{dd1})}{=}\sum_{j=1}^3  \frac{m_j^2 K_j}{2\sigma \sqrt{d}}
\stackrel{(\ref{d_1})}{=}
\frac{2\, \sigma d}{2\sigma \sqrt{d}}=\sqrt{d} \ .
$$
\qed

\begin{tests}
$$
\begin{array}{c|c|c|c|c}
   & P_1 & P_2 & P_3 & P_{\ast}   \\
   & m_1 & m_2 & m_3 & \sqrt{d}              \\
\hline
1. & (2,6) & (1,1) & (5,1) &  \left( \frac{1}{2866} \left( 4103+1833 \sqrt{15} \right)  \ , \  \frac{1}{8598} \left( 29523-4481 \sqrt{15} \right) \right)   \\
   &   2   &   3   &   4   &   \approx (3.9086, 1.4152)  \\
    &      &      &      & \sqrt{d}=2\sqrt{79+15\sqrt{15}} \approx 23.4174 \\
\hline
2. & (2,6) & (1,1) & (5,1) &  \left( \frac{751}{485}, \frac{647}{485} \right)   \\
   &   3   &   5   &   4   &   \approx (1.5484, 1.3340)   \\
  & & & &  \sqrt{d}= \sqrt{970} \approx 31.1448 \\
\hline
3. & (0,0) & (2,0) & (-\sqrt{2},\sqrt{2}) &  \left( 1-\frac{1}{\sqrt{2}}-\frac{3}{\sqrt{110}}  \ , \  \frac{1}{\sqrt{2}}-\frac{3}{\sqrt{55}}-\frac{3}{\sqrt{110}} \right)  \\
   &   3/2   &   2   &   2   &   \approx (0.0068, 0.0165)  \\
  & & & & \sqrt{d}= \sqrt{32+\frac{23}{\sqrt{2}}+3\sqrt{\frac{55}{2}}}   \approx 7.9997 \\
\hline
4. & (39,57) & (22,42) & (42,75) &    \\
   & 18 & 41 & 52 &    \\
   &     &      &       &   \approx (37.0432, 61.4053)   \\
  & & & &  \sqrt{d}=\sqrt{3068047+3915\sqrt{7511}}  \approx 1845.8994
\end{array}
$$
\end{tests}
The exact coordinates of $ P_{\ast} $ in Test 1.4  are as follows:
$$
\begin{array}{lcl}
x_{\ast} & =& \frac{296577529815837}{9297789607234}  +\frac{357441196078431}{6020318770684015} \sqrt{7511} \ , \\
  & & \\
y_{\ast} & =&
 \frac{271001243105952}{4648894803617} +\frac{432306390086253}{12040637541368030}\sqrt{7511}  \ .
\end{array}
$$


\section{Geometry}
\setcounter{equation}{0}
\setcounter{theorem}{0}
\setcounter{example}{0}

 Let us give an interpretation for some constants appeared in Theorem \ref{Th2}. Being rewritten in an alternative form, the constant (\ref{S})
$$
S= \left| \det \left(
\begin{array}{ccc}
1 & 1 & 1 \\
x_1 & x_2 & x_3 \\
y_1 & y_2 & y_3
\end{array}
\right)  \right|
$$
is recognized as the doubled area of the triangle $ P_1P_2P_3 $. As for the constant (\ref{sigma}), factorization of  the radicand in its right-hand side leads one to the form
$$
\sigma=
2 \sqrt{\frac{m_1+m_2+m_3}{2} \left( \frac{m_1+m_2+m_3}{2}-m_1 \right)\left( \frac{m_1+m_2+m_3}{2}-m_2 \right)\left( \frac{m_1+m_2+m_3}{2}-m_3 \right)}
$$
which can be treated as the Heron's formula for the doubled area of a triangle formed by the triple of weights $ m_1,m_2,m_3 $. Under the restrictions (\ref{r123}), such a triangle exists.
Construct now this triangle and denote its angles as shown in Figure 1.

\begin{figure}
\includegraphics{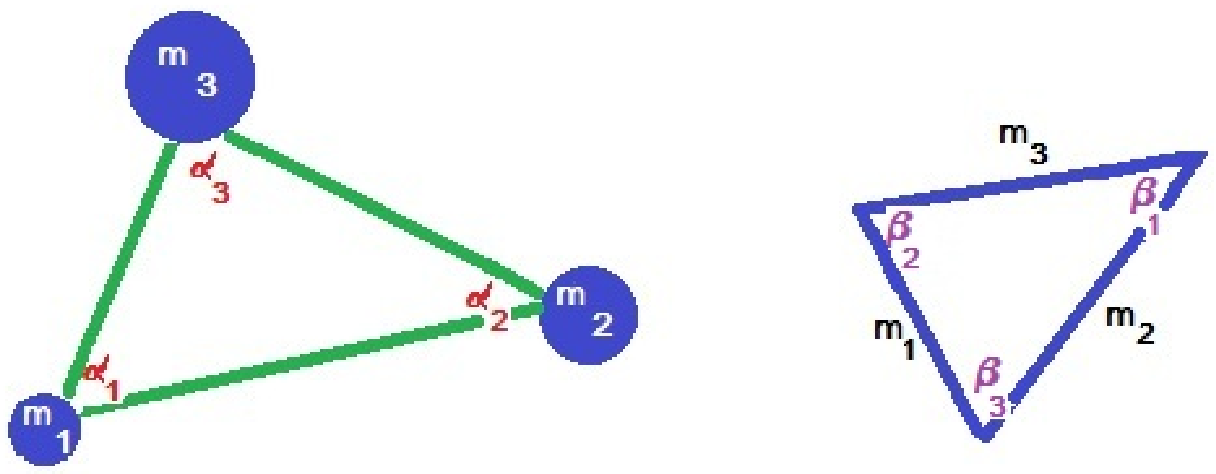}
\caption{}
\label{fig1}
\end{figure}

Then the first formula from (\ref{K1K2K3}) can be represented with the aid of the law of cosines as
$$
K_1=\sigma S \left( \frac{r_{12}^2+r_{13}^2-r_{23}^2}{S}+\frac{m_2^2+m_3^2-m_1^2}{\sigma}  \right)=
\sigma S \left( \frac{2\,r_{12}r_{13}\cos \alpha_1}{S}+\frac{2\,m_2m_3\cos \beta_1}{\sigma}  \right)
$$
$$
=
2\,\sigma S (\cot \alpha_1 + \cot \beta_1) \ .
$$

On rewriting the first condition from (\ref{r123}) in the form $ \cos \alpha_1 + \cos \beta_1 > 0 $, one can conclude that $ \cot \alpha_1 + \cot \beta_1>0 $ and, thus, $ K_1> 0 $.
In a similar way the expressions for $ K_2 $ and $ K_3 $ can be deduced, and established that under the restrictions (\ref{r123}) they both are positive. This completes the proof of Theorem \ref{Th2}.

\textbf{Remark.} Set the \emph{dual} generalized Fermat-Torricelli problem: let the triangle be composed of the sides with the lengths equal to $ m_1, m_2 $ and $ m_3 $; let the weights $ r_{12}, r_{23}, r_{13} $
be placed in its vertices as shown in Figure 2.

\noindent The minimum value for the objective function will be the same as in the direct problem since  (\ref{d_2}) is equivalent to
$$
2\, S \sigma + \frac{1}{2}\left[r_{12}^2(m_{1}^2+m_{2}^2-m_{3}^2)+
r_{13}^2(m_{1}^2+m_{3}^2-m_{2}^2)+r_{23}^2(m_{2}^2+m_{3}^2-m_{1}^2) \right] \ .
$$
\newpage
\begin{figure}
\includegraphics{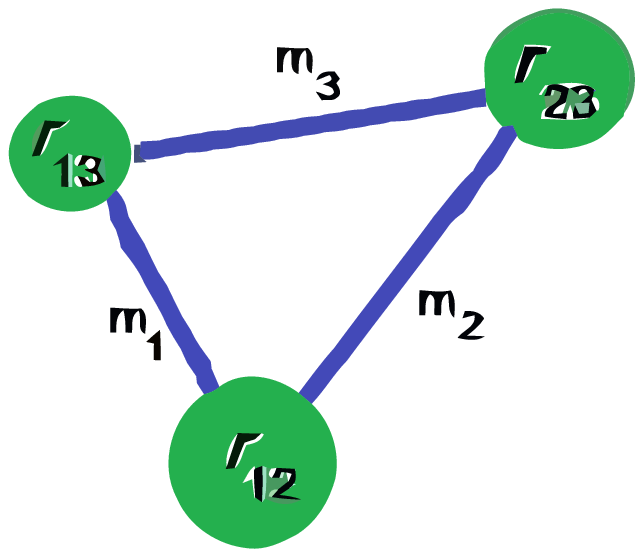}
\caption{}
\label{fig2}
\end{figure}

\section{Classical Fermat-Torricelli problem}
\setcounter{equation}{0}
\setcounter{theorem}{0}
\setcounter{example}{0}

Consider now the case $ m_1=m_2=m_3=1 $.

\begin{theorem} \label{Th3} Let all the angles of the triangle $ P_1P_2P_3 $ be less than $ 2\pi/3 $, or, equivalently:
$$ r_{12}^2+r_{13}^2+r_{12}r_{13}-r_{23}^2>0,\  r_{23}^2+r_{12}^2+r_{12}r_{23}-r_{13}^2>0,\  r_{13}^2+r_{23}^2+r_{13}r_{23}-r_{12}^2>0 \ . $$
 The coordinates of Fermat-Torricelli point for this triangle are as follows:
\begin{equation}
x_{\ast}=\frac{k_1k_2k_3}{2 \sqrt{3} S  d} \left(\frac{x_1}{k_1}+\frac{x_2}{k_2}+\frac{x_3}{k_3} \right), \
y_{\ast}=\frac{k_1k_2k_3}{2 \sqrt{3} S d} \left(\frac{y_1}{k_1}+\frac{y_2}{k_2}+\frac{y_3}{k_3} \right)
\label{FTp}
\end{equation}
with the corresponding minimum value of the objective function:
$$
F(x_{\ast},x_{\ast})= \min_{(x,y)} \sum_{j=1}^3 \sqrt{(x-x_j)^2+(y-y_j)^2} = \sqrt{d} \ .
$$
Here
\begin{equation}
d=\frac{1}{\sqrt{3}}(k_1+k_2+k_3)=\frac{r_{12}^2+r_{13}^2+r_{23}^2}{2}+ \sqrt{3}\, S
\label{FTv}
\end{equation}
and
$$
k_1=\frac{\sqrt{3}}{2}(r_{12}^2+r_{13}^2-r_{23}^2)+S , \ k_2=\frac{\sqrt{3}}{2}(r_{23}^2+r_{12}^2-r_{13}^2)+S , \
k_3=\frac{\sqrt{3}}{2}(r_{13}^2+r_{23}^2-r_{12}^2)+S .
$$
with the rest of the parameters coinciding with those from Theorem \ref{Th2}.
\end{theorem}

It turns out that expressions  (\ref{FTp}), being represented as rational fractions with respect to $ \{x_j,y_j\}_{j=1}^3 $, can be reduced further to the form where denominators
become "area free" \cite{b6}:

\textbf{Corollary.}  Under the conditions of Theorem \ref{Th3}, the coordinates of Fermat-Torricelli point are as follows:
\begin{eqnarray}
x_{\ast}&=& \frac{1}{2 \sqrt{3}\, d} \big[ (x_1+x_2+x_3) | \tilde S | + \sqrt{3} \left(x_1r_{23}^2+x_2r_{13}^2+x_3r_{12}^2 \right)    \label{FTp1x} \\
        & &  +3\, \sgn ( \tilde S) \left|
\begin{array}{ccc}
1 & 1 & 1 \\
y_1 & y_2 & y_3 \\
x_2x_3+y_2y_3 & x_1x_3+y_1y_3 & x_1x_2+y_1y_2
\end{array}
\right|   \big], \nonumber \\
y_{\ast}&=& \frac{1}{2 \sqrt{3}\, d} \big[ (y_1+y_2+y_3) | \tilde S | + \sqrt{3} \left(y_1r_{23}^2+y_2r_{13}^2+y_3r_{12}^2 \right)    \label{FTp2x} \\
        & &  -3\, \sgn (\tilde S) \left|
\begin{array}{ccc}
1 & 1 & 1 \\
x_1 & x_2 & x_3 \\
x_2x_3+y_2y_3 & x_1x_3+y_1y_3 & x_1x_2+y_1y_2
\end{array}
\right|  \big] \ . \nonumber
\end{eqnarray}
Here
$$
\tilde S=  \det \left(
\begin{array}{ccc}
1 & 1 & 1 \\
x_1 & x_2 & x_3 \\
y_1 & y_2 & y_3
\end{array}
\right) \ .
$$

\textbf{Remark.} Result of the last corollary can be extended to the generalized Fermat-Torricelli problem: numerators and denominators of
the formulae (\ref{GFTp}) can be reduced by the common factor $ S $. We do not present here the resulting expressions since they are inelegantly
ponderous \cite{b6}.

\begin{tests}
$$
\begin{array}{c|c|c|c|c|c|c}
   & P_1 & P_2 & P_3 & x_{\ast} & y_{\ast} & \sqrt{d}   \\
 \hline
1. & (1,1) & (3,5) & (7,2) &  \frac{2}{687} \left( 1029 + 79 \sqrt{3}\right)  & \frac{1}{687} \left(  1053 + 647 \sqrt{3} \right) & \sqrt{41+22\sqrt{3}}  \\
   &     &      &      &   \approx 3.3939 & \approx 3.1639 & \approx 8.8941 \\
\hline
2. & (1,2) & (3,3) & (4,1) &   \frac{1}{6}(15+\sqrt{3})  & \frac{1}{2}(3+\sqrt{3})  & \sqrt{10+5\sqrt{3}}  \\
   &     &     &     &   \approx 2.7886 & \approx 2.3660 & \approx 4.3197 \\
\hline
3. & (0,0) & (399,0) & (\frac{5005}{38},\ \frac{9555\sqrt{3}}{38}) & \frac{21255}{133}    & \frac{8580\sqrt{3}}{133} & 784  \\
   &      &      &      &   \approx 159.8120 & \approx 111.7368 &   \\
\hline
4. &  (0,0) & (2,0) & (0,1) &  \frac{1}{13} +\frac{4}{39}\sqrt{3}  & \frac{8}{13} -\frac{7}{39}\sqrt{3} &  \sqrt{5+2\sqrt{3}} \\
   &      &      &  & \approx  0.2545  & \approx 0.3045  & \approx 2.9093  \\
\end{array}
$$
\end{tests}

Test 2.3 is generated from  \cite{b8}, test 2.4 is taken from \cite{b9}.

\section{Inverse problem}
\setcounter{equation}{0}
\setcounter{theorem}{0}
\setcounter{example}{0}

Given the coordinates of the point $ P_{\ast} =(x_{\ast},y_{\ast}) $ lying inside the triangle $ P_1P_2P_3 $, find the values for the weights $ m_1,m_2,m_3 $
with the aim for the corresponding function $ F(x,y) $ to posses a minimum point precisely at $ P_{\ast} $.

\begin{theorem} \label{Th4} Let the vertices of the triangle $ P_1P_2P_3 $ be counted counterclockwise. Then for the choice
\begin{equation}
m_1^{\ast} = |P_{\ast}P_1| \cdot \left|
\begin{array}{ccc}
1 & 1 & 1 \\
x_{\ast} & x_2 & x_3 \\
y_{\ast} & y_2 & y_3
\end{array}
\right|, \
m_2^{\ast} = |P_{\ast}P_2| \cdot \left|
\begin{array}{ccc}
1 & 1 & 1 \\
x_1 & x_{\ast} & x_3 \\
y_1 & y_{\ast} & y_3
\end{array}
\right|,\
m_3^{\ast} = |P_{\ast}P_3| \cdot \left|
\begin{array}{ccc}
1 & 1 & 1 \\
x_1 & x_2 & x_{\ast} \\
y_1 & y_2 & y_{\ast}
\end{array}
\right|
\label{masses}
\end{equation}
the function
$$ F(x,y) = \sum_{j=1}^3 m_{j}^{\ast}\sqrt{(x-x_j)^2+(y-y_j)^2} $$
has its stationary point at $ P_{\ast} $.  Provided that the latter is chosen inside the triangle
$P_1P_2P_3$ the values (\ref{masses}) are all positive and
\begin{equation}
F(x_{\ast},y_{\ast})=\min_{(x,y)} F(x,y)=
\left|
\begin{array}{cccc}
1 & 1 & 1  & 1\\
 x_{\ast} & x_1 & x_2 & x_3  \\
 y_{\ast} & y_1 & y_2 & y_3  \\
x_{\ast}^2+y_{\ast}^2 & x_1^2+y_1^2 & x_2^2+y_2^2 & x_3^2+y_3^2
\end{array}
\right| \ .
\label{value}
\end{equation}
\end{theorem}

\textbf{Proof}. Substitute $ x=x_{\ast},y=y_{\ast} $ and the values (\ref{masses}) into the left-hand side of the first equation from (\ref{sys1}):
\begin{equation}
(x_{\ast}-x_1)\left|
\begin{array}{ccc}
1 & 1 & 1 \\
x_{\ast} & x_2 & x_3 \\
y_{\ast} & y_2 & y_3
\end{array}
\right|+
(x_{\ast}-x_2)
\left|
\begin{array}{ccc}
1 & 1 & 1 \\
x_1 & x_{\ast} & x_3 \\
y_1 & y_{\ast} & y_3
\end{array}
\right|+
(x_{\ast}-x_3)
\left|
\begin{array}{ccc}
1 & 1 & 1 \\
x_1 & x_2 & x_{\ast} \\
y_1 & y_2 & y_{\ast}
\end{array}
\right| \ .
\label{lc1}
\end{equation}
Represent this combination of the third order determinants in the form of the fourth order determinant, namely
$$
\left|
\begin{array}{cccc}
1 & 1 & 1  & 1\\
 x_{\ast} & x_1 & x_2 & x_3  \\
 y_{\ast} & y_1 & y_2 & y_3  \\
0 & x_{\ast}-x_1 & x_{\ast}-x_2 & x_{\ast}-x_3
\end{array}
\right|
$$
(expansion by its last row coincides with (\ref{lc1})). Now add the second row to the last one:
$$
\left|
\begin{array}{cccc}
1 & 1 & 1  & 1\\
 x_{\ast} & x_1 & x_2 & x_3  \\
 y_{\ast} & y_1 & y_2 & y_3  \\
 x_{\ast} & x_{\ast} & x_{\ast} & x_{\ast}
\end{array}
\right| \ .
$$
In this determinant the first row is proportional to the last one; therefore the determinant equals just zero.
The second equality from (\ref{sys1}) can be verified in a similar manner.

Let us evaluate $ F(x_{\ast},y_{\ast}) $:
\begin{eqnarray*}
F(x_{\ast},y_{\ast}) & = & \left[(x_{\ast}-x_1)^2+(y_{\ast}-y_1)^2 \right] \left|
\begin{array}{ccc}
1 & 1 & 1 \\
x_{\ast} & x_2 & x_3 \\
y_{\ast} & y_2 & y_3
\end{array}
\right| \\
& + & \left[(x_{\ast}-x_2)^2+(y_{\ast}-y_2)^2 \right]
\left|
\begin{array}{ccc}
1 & 1 & 1 \\
x_1 & x_{\ast} & x_3 \\
y_1 & y_{\ast} & y_3
\end{array}
\right| \\
& + & \left[(x_{\ast}-x_3)^2+(y_{\ast}-y_3)^2 \right]
\left|
\begin{array}{ccc}
1 & 1 & 1 \\
x_1 & x_2 & x_{\ast} \\
y_1 & y_2 & y_{\ast}
\end{array}
\right| \ .
\end{eqnarray*}
To prove the equality (\ref{value}) let us split it into the $ x $-part and the $ y $-part. First keep the $ x $-terms in brackets of the previous formula:
$$
(x_{\ast}-x_1)^2 \left|
\begin{array}{ccc}
1 & 1 & 1 \\
x_{\ast} & x_2 & x_3 \\
y_{\ast} & y_2 & y_3
\end{array}
\right|
+  (x_{\ast}-x_2)^2
\left|
\begin{array}{ccc}
1 & 1 & 1 \\
x_1 & x_{\ast} & x_3 \\
y_1 & y_{\ast} & y_3
\end{array}
\right|
+  (x_{\ast}-x_3)^2
\left|
\begin{array}{ccc}
1 & 1 & 1 \\
x_1 & x_2 & x_{\ast} \\
y_1 & y_2 & y_{\ast}
\end{array}
\right| \ .
$$
Similar to the proof of the first part of the theorem, represent this linear combination as the determinant of the fourth order
$$
=
\left|
\begin{array}{cccc}
1 & 1 & 1  & 1\\
 x_{\ast} & x_1 & x_2 & x_3  \\
 y_{\ast} & y_1 & y_2 & y_3  \\
 0 & (x_{\ast}-x_1)^2 & (x_{\ast}-x_2)^2 & (x_{\ast}-x_3)^2
\end{array}
\right| \ .
$$
Multiply the first row by $ (-x_{\ast}^2) $,  the second one by $ 2\, x_{\ast} $ and add the obtained rows to the last one:
\begin{equation}
=
\left|
\begin{array}{cccc}
1 & 1 & 1  & 1\\
 x_{\ast} & x_1 & x_2 & x_3  \\
 y_{\ast} & y_1 & y_2 & y_3  \\
x_{\ast}^2  & x_1^2 & x_2^2 & x_3^2
\end{array}
\right| \ .
\label{valuex}
\end{equation}
The $ y $-part of the equality (\ref{value}) can be proved in exactly the same manner with the resulting determinant differing from (\ref{valuex}) only in its
last row. The linear property of the determinant with respect to its rows completes the proof of (\ref{value}). \qed

\textbf{Remark.} Solution of the inverse problem is determined up to a common positive multiplier, i.e. the solution triple $ (m_1,m_2,m_3) $ is defined by the
value of the ratio $ m_1 : m_2 : m_3 $.  Up to this remark\footnote{In the language of railway transportation, it is equivalent to the fact that the optimal position
of junction is independent of the currency of the state: it does not matter whether in roubles or in dollars the freight charge is nominated.},
solution of the inverse problem is unique:  we have proved this statement via direct computations starting from the formulae (\ref{GFTp}).

\begin{example}  Let $ P_1=(2,6),P_2=(1,1),P_3=(5,1) $ and
$$ P_{\ast}=\left(\frac{1}{2866} \left( 4103+1833 \sqrt{15} \right) ,\ \frac{1}{8598} \left( 29523-4481 \sqrt{15} \right) \right) \ . $$
Find the values for the weights $ m_1^{\ast}, m_2^{\ast}, m_3^{\ast} $ from Theorem \ref{Th4}.
\end{example}

\textbf{Solution.} Formulae (\ref{masses}) give:
\begin{eqnarray*}
m_1^{\ast}&=&\frac{2(20925-4481\sqrt{15})}{18481401}\sqrt{316380606+35999826\sqrt{15}},\\
m_2^{\ast}&=&\frac{2(15105-2342\sqrt{15})}{6160467}\sqrt{75400161-9169767\sqrt{15}},\\
m_3^{\ast}&=&\frac{8(-1185+15988\sqrt{15})}{18481401}\sqrt{8335761-2050623\sqrt{15}},
\end{eqnarray*}
with
$$
F(x_{\ast},y_{\ast})=\frac{1}{4299}(-333980+193436\sqrt{15}) \ .
$$
Now compare the obtained result with the one represented in Test 1.1: according with the last remark one might expect that
$$ m_1^{\ast} : m_2^{\ast} : m_3^{\ast} = 2 : 3 : 4 \ . $$
Let us verify this fact:
$$
\frac{m_1^{\ast}}{m_2^{\ast}}=\frac{1}{3}\cdot \frac{20925-4481\sqrt{15}}{15105-2342\sqrt{15}} \cdot \sqrt{\frac{316380606+35999826\sqrt{15}}{75400161-9169767\sqrt{15}}}
$$
$$
=\frac{1}{3} \cdot \frac{(20925-4481\sqrt{15})(15105+2342\sqrt{15})}{145886565}
$$
$$
 \times \sqrt{\frac{(316380606+35999826\sqrt{15})(75400161+9169767\sqrt{15})}{4423914876311586}}
$$
$$
=\frac{1}{3}\cdot \frac{671-79\sqrt{15}}{617} \cdot \sqrt{\frac{543856+106018\sqrt{15}}{83521}}=\frac{1}{3}\cdot \frac{671-79\sqrt{15}}{617} \cdot \frac{671+79\sqrt{15}}{289}=\frac{2}{3} \ .
$$
\qed

Let us discuss now  geometrical meaning of the constants from Theorem \ref{Th4}. The value $ m_1^{\ast} $ equals the doubled product of the distance from $ P_1 $ to $P_{\ast} $  by the area of the triangle $ P_{\ast}P_2P_3 $.

\begin{figure}
\includegraphics{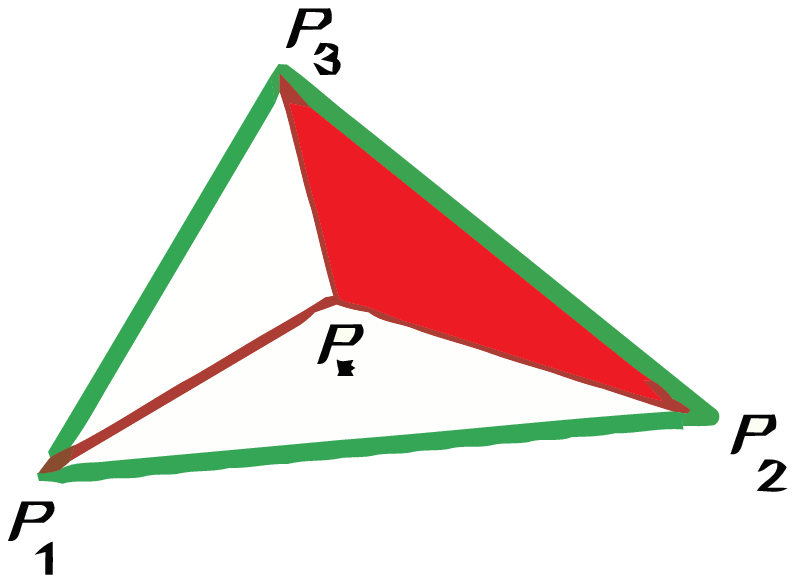}
\caption{}
\label{fig3}
\end{figure}

The first statement of the theorem is equivalent to the fact that
$$ \overrightarrow{P_{\ast}P_1} \cdot S_{_{\triangle P_{\ast}P_2P_3}}+\overrightarrow{P_{\ast}P_2} \cdot S_{_{\triangle P_{\ast}P_3P_1}}+\overrightarrow{P_{\ast}P_3} \cdot S_{_{\triangle P_{\ast}P_1P_2}}=\overrightarrow{\mathbb O} \ . $$

Finally, the constant (\ref{value}) is connected with the following one
$$
h=-\frac{1}{S} \left|
\begin{array}{cccc}
1 & 1 & 1  & 1\\
 x_{\ast} & x_1 & x_2 & x_3  \\
 y_{\ast} & y_1 & y_2 & y_3  \\
x_{\ast}^2+y_{\ast}^2 & x_1^2+y_1^2 & x_2^2+y_2^2 & x_3^2+y_3^2
\end{array}
\right| \ ,
$$
which is known \cite{b7} as \emph{the power of the point} $ P_{\ast} $ \emph{with respect to the circle} through the points $ P_1, P_2 $ and $ P_3 $ (circumscribed circle of the triangle).
If one denotes by $ C $ the circumcenter of the triangle $ P_1P_2P_3 $ then
\begin{equation}
 h=|CP_{\ast}|^2-|CP_j|^2
 \label{power}
\end{equation}
and provided that  $ P_{\ast} $ lies inside this triangle, this value is negative.

Results of the present section can evidently be extended to the case of three (and more) dimensions:

\begin{theorem} \label{Th5} Let the points $ \{P_j=(x_j,y_j,z_j)\}_{j=1}^4 $ be noncoplanar, and, in addition, be counted in such a manner that the value
of the determinant
\begin{equation}
V=
\left|
\begin{array}{cccc}
1 & 1 & 1  & 1\\
 x_{1} & x_2 & x_3 & x_4  \\
 y_{1} & y_2 & y_3 & y_4  \\
 z_{1} & z_2 & z_3 & z_4
\end{array}
\right|
\label{V}
\end{equation}
is positive. Then for the choice
\begin{equation}
\left\{m_j^{\ast} = |P_{\ast}P_j| \cdot V_j \right\}_{j=1}^4
\label{masses3d}
\end{equation}
where $ V_j $ equals the determinant obtained on replacing the $ j $-th column of (\ref{V}) by the column\footnote{Here $ \mbox{}^{\top} $ denotes transposition.} $ \left[1,x_{\ast},y_{\ast},z_{\ast} \right]^{\top} $,
the function
$$ F(x,y,z) = \sum_{j=1}^4 m_{j}^{\ast}\sqrt{(x-x_j)^2+(y-y_j)^2+(z-z_j)^2} $$
has its stationary point at $ P_{\ast}=(x_{\ast},y_{\ast},z_{\ast}) $.  If $ P_{\ast} $ lies inside the tetrahedron $ P_1P_2P_3P_4 $ then
the values (\ref{masses3d}) are all positive and
$$
F(x_{\ast},y_{\ast},z_{\ast})=\min_{(x,y,z)} F(x,y,z)
$$
\begin{equation}
=-
\left|
\begin{array}{ccccc}
1 & 1 & 1  & 1 & 1 \\
 x_{\ast} & x_1 & x_2 & x_3 & x_4  \\
 y_{\ast} & y_1 & y_2 & y_3  & y_4 \\
z_{\ast} & z_1 & z_2 & z_3  & z_4 \\
x_{\ast}^2+y_{\ast}^2+z_{\ast}^2 & x_1^2+y_1^2+z_1^2 & x_2^2+y_2^2+z_2^2 & x_3^2+y_3^2+z_3^2 & x_4^2+y_4^2+z_4^2
\end{array}
\right| \ .
\label{value3d}
\end{equation}
\end{theorem}

Geometrical meanings of the values appeared in the last theorem are similar to their counterparts from Theorem \ref{Th4}.
For instance, the value (\ref{V}) equals six times the volume of tetrahedron  $ P_1P_2P_3P_4 $, while the value (\ref{value3d}) divided by $ V $ is known \cite{b7} as \emph{the
power of the point }$ P_{\ast} $ \emph{with respect to a sphere} circumscribed to that tetrahedron; it is equivalent to (\ref{power}) where $ C $ this time stands for the circumcenter of the tetrahedron.

\section{Conclusions}

Analytical solution for the generalized Fermat-Torricelli problem is presented. The three point case is completely solved in "extended
radicals": in addition to elementary and extraction of roots operations the sign function is utilized in the formulae. For further
investigations remains the treatment of the multidimensional $ n $ point case, although some theoretical results like \cite{b12}  do not give reason to
hope for the nice (e.g. extended radicals) solution.

\setcounter{equation}{0}
\setcounter{theorem}{0}
\setcounter{example}{0}

\end{document}